\documentclass{svjour3}
\smartqed
\usepackage{amssymb}
\usepackage{amsmath}
\usepackage{graphicx}
\usepackage{subfigure}
\usepackage{setspace}
\usepackage{multirow}

\journalname{Quantum Information Processing}
\begin{document}
\title{Security of a single-state semi-quantum key distribution protocol }

\author{Wei Zhang \and Daowen Qiu \and Paulo Mateus}

\institute{Wei Zhang \and Daowen Qiu \at
              School of Computer Science, Sun Yat-sen University, Guangzhou 510006, China \\
              The Guangdong Key Laboratory of Information Security Technology, Sun Yat-sen
University, Guangzhou 510006, China\\
              \email{issqdw@mail.sysu.edu.cn}\\
\and
Wei Zhang  \at
              School of Mathematics and Statistics, Qiannan Normal College for Nationalities, Duyun 558000, China
              \and Paulo Mateus \at
 SQIG--Instituto de Telecomunica\c{c}\~{o}es, Departamento de Matem\'{a}tica, Instituto Superior T\'{e}cnico, Av. Rovisco Pais
              1049-001, Lisbon, Portugal}

\maketitle
\begin{abstract}
Semi-quantum key distribution protocols are allowed to set up a secure secret key between two users. Compared with their full quantum counterparts, one of the two users is restricted to perform some ``classical" or ``semi-quantum" operations, which makes them potentially easily realizable by using less quantum resource. However, the semi-quantum key distribution protocols mainly rely on a two-way quantum channel. The eavesdropper has two opportunities to intercept the quantum states transmitted in the quantum communication stage. It may allow the eavesdropper to get more information and make the security analysis more complicated. In the past ten years, many semi-quantum key distribution protocols have been proposed and proved to be robust. However, there are few works concerned about their unconditional security. It is doubted that how secure the semi-quantum ones are and how much noise they can tolerate to establish a secure secret key. In this paper, we prove the unconditional security of a single-state semi-quantum key distribution protocol proposed by $Zou$ et al. in [Phys. Rev. A. 79]. We present a complete proof from information theory aspect by deriving a lower bound of the protocol's key rate in the asymptotic scenario. Using this bound, we figure out an error threshold value such that all error rates that are less than this threshold value, the secure secret key can be established between the legitimate users definitely. Otherwise, the users should abort the protocol. We make an illustration of the protocol under the circumstance that the reverse quantum channel is a depolarizing one with parameter $q$. Additionally, we compare the error threshold value with some full quantum protocols and several existing semi-quantum ones whose unconditional security proofs have been provided recently.
\keywords{Semi-quantum key distribution protocol \and Von Neumann entropy \and Secret key rate \and Error rate \and Measurement}

\end{abstract}

\section{Introduction}

Quantum key distribution (QKD) protocols are designed to share secret keys between two legitimate users. One is called the sender Alice and the other is the receiver Bob. The shared keys can secure against all powerful quantum adversary, even it is powerful enough to perform any operators allowed in quantum physics. In 1984, $Bennett$ and $Brassard$ \cite{1} proposed a protocol to share secret keys  by using quantum method, which is called the BB84 protocol. After that, numerous QKD protocols have been developed including B92 \cite{2}, SARG 04 \cite{3}, three states BB84 \cite{4} and so on. Furthermore, some QKD protocols have been proved to be unconditionally secure \cite{5,6,7}. In classical cryptography, only the one-time pad encryption has been proved to have unconditional security. Then we can obtain unconditionally secure encryption methods by combining QKD with one-time pad.

QKD protocols have showed us that we can share unconditional secure secret keys between quantum users. It is of great interest to achieve this goal by using as few ``quantum resource" as possible. In 2007, $Boyer$ et al. \cite{8} designed the first semi-quantum key distribution (SQKD) protocol, which can share secure secret keys between quantum Alice and classical Bob (BKM2007). Here we say Bob is classical as he is limited to do some classical performance and has no quantum computational ability. Since their idea of genius, numerous SQKD protocols have been developed by scholars \cite{9,10,11,12,13,14,15,16,17,18,19}. For instance, $Zou$ et al. \cite{11} proposed several SQKD protocols with less than four quantum states based on BKM2007. In Ref. \cite{12}, a multi-user protocol was developed which can make a quantum user $A$ share secure secret keys with several classical users $B,C$ and so on. A SQKD protocol was proposed based on quantum entanglement in Ref. \cite{13}. Recently, $Zou$ et al. \cite{17} have presented a new SQKD protocol, in which the classical Bob has no need to measure any quantum state. $Krawec$ \cite{19} designed a new SQKD protocol with one quantum state, in which the receiver Bob's reflection can contribute to the raw key as well.

In an SQKD protocol, the quantum user Alice sends a quantum state prepared in arbitrary basis to the other classical user Bob, who is restricted to performing one of the two operations:
\begin{enumerate}

\item  Bob chooses to measure and resend (SIFT). He measures the state he received in the computational basis $Z=\{|0\rangle, |1\rangle\}$ and resends the result state to Alice. In other words, Bob sends the state $|0\rangle$ to Alice when he gets the measurement outcome $0$. Otherwise, he sends $|1\rangle$ to Alice.

\item  Bob chooses to reflect (CTRL). He just makes the state pass through his own lab undisturbed and returns it to Alice. Under this circumstance, Bob gets nothing about the state.
\end {enumerate}
Then Alice chooses to measure the state in arbitrary basis randomly when she receives it. From the above, we can see Bob resends a quantum state in both of the two case regardless of his choice. That is to say, there are two-way quantum communication channel in an SQKD protocols. The forward channel is from the sender Alice to the receiver Bob, the reverse one is from Bob to Alice.

Due to the reliance on a two-way quantum communication channel, the security analysis of SQKD protocols is more difficult than their full quantum counterparts because the eavesdropper Eve can attack the quantum state twice in the quantum communication stage. This may greatly increase the possibility for Eve to gain more information on A or B's raw key and make the security analysis more complicated. Therefore, most existing SQKD protocols are limited to be proved robust rather than unconditionally secure. We say a protocol is robust if any attacker can get nontrivial information on the $A$ or $B$'s secret keys, his dishonest behavior will be detected by the legitimate users $A$ and $B$ with nonzero probability \cite{10}. Then the robustness of SQKD protocols can only assure any attack can be detected, but it cannot tell us how much noise the protocol can tolerate to distill a secure key after applying the technique of error correction and privacy amplification.

Recently, the unconditional security of SQKD protocols has  attracted more and more scholars' attention. Ref. \cite{20} studied the relationship between the disturbance and the amount of information gained by Eve under the assumption that Eve just performs individual attacks during the whole quantum communication stage. $Krawec$ \cite{21} showed that all Eve's collective attack strategies are equivalent to a restricted attack operation when Alice just sends the same quantum state that is known in public in each iteration. In Ref. \cite{22,23}, $Krawec$ proved the unconditional security of BKM2007 by deriving an expression for the key rate as a function of the quantum channel noise which can be observed by the legitimate users in the asymptotic scenario. To the best of our knowledge, this is the first unconditional security proof of an SQKD protocol. Then  $Krawec$ \cite{16} proved the unconditional security of a mediated SQKD protocol allowing two semi-quantum users to share secure secret keys with the help of a quantum server. It is pointed out that the unconditional security can be promised even under the circumstance that the quantum server is an all-powerful adversary. After that, $Krawec$ \cite{19} provided an unconditional security proof of an SQKD protocol with one quantum state, in which the classical Bob's reflection contributes to the raw key as well.

Though some initiate works have been done by some geniuses, there are still various SQKD protocols that need to be concerned about their unconditional security. In this paper, we utilize $Krawec$'s methods first presented in Ref. \cite{21,22} to analyze the unconditional security of an SQKD protocol with one quantum state proposed by $Zou$ et al. in Ref. \cite{11}. We also derive an expression for the key rate as a function of the quantum channel parameters in the asymptotic scenario. Then we further derive a lower bound of the key rate. Using this bound, we figure out an error threshold value such that for all error rates less than this threshold value, the secure secret key can be established between the legitimate users definitely. Otherwise, the users should abort the protocol. We make an illustration of the protocol under the circumstance that the reverse quantum channel is a depolarizing one with parameter $q$. Additionally, we compare the error threshold value with some full quantum protocols and several existing semi-quantum ones whose unconditional security proofs have been provided recently.

The rest of this paper is organized as follows. First, in Section 2 we give some preliminaries, including some notations and the single-state SQKD protocol proposed in the Ref. \cite{11}. Then we  present the whole unconditional security proof in detail in Section 3. After that, in Section 4 we take an example to make an illustration by considering the protocol under the circumstance that the reverse quantum channel is a depolarizing one with parameter $q$. Finally, Section 5 makes a short conclusion and gives some future issues.

\section {Preliminaries}
\subsection{Notations\cite{24,25}}

In the interest of readability, we first present some notations which will appear in the next of this paper.

The computational basis denoted as $Z$ basis is the two state set $\{|0\rangle, |1\rangle\}$, the $Hadamard$ basis denoted as $X$ basis is the set $\{|+\rangle, |-\rangle\}$, where
\begin{eqnarray}
&& |+\rangle = \frac{|0\rangle+|1\rangle}{\sqrt{2}},\\
&& |-\rangle = \frac{|0\rangle-|1\rangle}{\sqrt{2}}.
\end{eqnarray}
Given a number $z\in\mathbb{C}$, we denote $Re(z)$ and $Im(z)$ as its real and imaginary components respectively. If $U$ is a complex matrix (operator), its conjugate transpose (conjugate) denoted as $U^{*}$.

Consider a random variable $X$ and suppose each realization $x$ of $X$ belongs to the set $N=\{1,2,\cdots,i,\cdots,n\}$. Let $P_{X}(i)$ be the probability distribution of $X$,  then the Shannon entropy of $X$ is
\begin{eqnarray}
&& H(X)=H(P_{X}(1),\cdots,P_{X}(i),\cdots,P_{X}(n))=-\sum^{n}_{i=1}P_{X}(i)\log_{2}(P_{X}(i)).
\end{eqnarray}
Here we give the extra definition $0\log_{2}0=0$. If $N=2$, then $H(X)=h(P_{X}(1))$, where $h(x)=H(x,1-x)$ is the Shannon binary entropy function.

Let $\rho$ be a density operator acting on some finite dimensional Hilbert space $\mathcal{H}$ satisfying
\begin{eqnarray}
&& \rho =\sum^{n}_{i=1}\lambda_{i}|v_{i}\rangle\langle v_{i}|,
\end{eqnarray}
where $\lambda_{i}(i=1,2,\cdots,n$) is the $i$-th eigenvalue of $\rho$ and $\{|v_{1}\rangle, |v_{2}\rangle,\cdots, |v_{n}\rangle\}$ is an orthonormal basis of $\mathcal{H}$. Then we denote $S(\rho)$ as its von Neumann entropy such that
\begin{eqnarray}
&& S(\rho) =H(\{\lambda_{i}\}_{i})=-\sum^{n}_{i=1}\lambda_{i}\log_{2}\lambda_{i}.
\end{eqnarray}
Note that if $\rho$ is a pure state, then $S(\rho)=0$ and $S(\rho)\geq 0$  holds for arbitrary state $\rho$.

Let $\rho$ be a classical quantum  state expressed as
\begin{eqnarray}
&& \rho =\sum^{n}_{i=1}P_{X}(i)|v_{i}\rangle\langle v_{i}|\otimes \rho_{i}.
\end{eqnarray}
Then
\begin{eqnarray}
&& S(\rho) =H(P_{X}(i))+\sum_{i=1}^{n} P_{X}(i)S(\rho_{i}) .
\end{eqnarray}

If $\rho_{AB}$ is a density operator acting on the bipartite space $\mathcal{H}_{A}\otimes \mathcal{H}_{B}$, we use $S(AB)$ to denote the von Neumann entropy of $\rho_{AB}$ and $S(B)$ the von Neumann entropy of $\rho_{B}$ where $S(B)=S(tr_{A}(\rho_{AB}))$. We use $S(A|B)$ to denote the von Neumann entropy of $A$'s system conditioned by system $B$ such that
\begin{eqnarray}
&& S(A|B) =S(AB)-S(B)=S(\rho_{AB})-S(tr_{A}(\rho_{AB})).
\end{eqnarray}

Let $n$ be the size of $A$ and $B$'s raw key of an SQKD protocol, $\ell(n)< n$ denotes the size of secure secret key distilled after error correction and privacy amplification. Let $r$ denote the key rate in the asymptotic scenario ($n \rightarrow \infty$), then
\begin{eqnarray}
&& r=\lim_{n\rightarrow \infty}\frac{\ell(n)}{n}\geq\inf (S(B|E)-H(B|A)),
\end{eqnarray}
where $H(B|A)$ is the conditional Shannon entropy and the infimum is over all collective attacks an eavesdropper Eve can perform. This is the Devetak-Winter key rate using reverse reconciliation as opposed to the usual direct reconciliation $(S(A|E)-H(A|B))$ \cite{19,26,27,28}.

\subsection{The protocol}

In this part, we briefly review the single-state SQKD protocol proposed in Ref. \cite{11}. In this protocol, the receiver Bob is limited to be semi-quantum or classical. The protocol consists of the following steps:

\begin{enumerate}
\item Alice prepares and sends $N$ quantum states $|+\rangle$ to Bob one by one, where $N=8n(1+\delta)$, $n$ is the desired length of the raw key, and $\delta > 0$ is a fixed parameter. Alice sends a quantum state only after receiving the previous one.
\item Bob generates a random string $b\in\{0,1\}^{N}$. When the $i$-th quantum state arriving, he chooses to CTRL it if $b_{i}=0$ or SIFT it if $b_{i}=1$.
\item Alice generates a random string $c\in \{0,1\}^{N}$. She measures the $i$-th quantum state in the $Z$ basis if $c_{i}=0$ and measures the $i$-th quantum state in the $X$ basis if $c_{i}=1$.
\item Alice announces $c$ and Bob announces $b$. They check the number of SIFT-Z bits (classical bits are produced by the process that Bob chooses to SIFT and Alice measures in the $Z$ basis). They abort the protocol if the number of SIFT-Z bits is less than $2n$.
\item Alice checks the error rate on the CTRL-$X$ bits. She and Bob abort the protocol if the error rate is higher than the predefined threshold $P_{t}$.
\item Alice chooses at random $n$ measure result of SIFT-$Z$ bits to be test bits. Alice and Bob check the error rate on the test bits. They abort the protocol if the error rate is higher than $P_{t}$.
\item Alice and Bob select the first $n$ remaining measure results of SIFT-$Z$ bits to be used as raw key bits.
\item Alice announces ECC(error correction code) and PA(privacy amplification) data, she and Bob use them to extract the $\ell(n)$-bit final key from the $n$-bit raw key.
\end{enumerate}

Note that, Alice prepares and sends only one quantum state $|+\rangle$ to Bob in each iteration of the protocol. The classical Bob chooses to CTRL or SIFT the state and Alice chooses to perform a measurement on it in $Z$ or $X$ basis randomly. When Bob chooses to SIFT and Alice chooses to measure in the $Z$ basis, they share a bit, in other words, only the SIFT-$Z$ bits contribute to the raw keys.

Note that, to enhance the protocol's efficiency, we can make Alice choose to measure in $Z$ basis and Bob choose to SIFT with greater probability, which was adapted in Ref. \cite {29}.

\section{Security proof}

 Three kinds of attack strategies are mainly talked about when analyzing the security of QKD protocols, including individual attack, collective attack and general attack (coherent attack/joint attack). Individual attack is a kind of attack strategy that Eve performs the same operation in each iteration of the protocol and measures her ancilla immediately; Collective attack is a typical attack strategy that Eve performs the same operation in each iteration of the protocol, but she can postpone to measure her ancilla in any future time; General attack is a kind of powerful attack strategy that Eve can perform any operation allowed by the laws of quantum physics in each iteration and postpone her measurement on her ancilla all by herself \cite{30}. Here we concentrate on Eve's collective attack at first.
\subsection{Modeling the protocol}

We use $\mathcal{H}_{A}$, $\mathcal{H}_{B}$ and $\mathcal{H}_{E}$ to denote Alice, Bob and Eve's Hilbert space respectively. $\mathcal{H}_{T}$ is the Hilbert space of the transit state. Generally, they are all assumed to be finite. We can use the density operator $\rho_{ABE}=\bigotimes^{N}_{i=1}\rho^{i}_{ABE}$ acting on the  Hilbert space $\bigotimes^{N}(\mathcal{H}_{A}\otimes \mathcal{H}_{B} \otimes \mathcal{H}_{E})$ to denote the mixed state of Alice, Bob and Eve during the whole quantum communication stage by applying the finite version of quantum de Finetti representation theorm \cite{31}, where $\rho^{i}_{ABE}$ models the joint system of each iteration of this protocol. As it is just restricted to the circumstance of collective attack, we just need to take one iteration for example to prove the unconditional security.

$Krawec$ has pointed out any collective attack $(U_{F}, U_{R})$ is equivalent to a restriction operation $(b, U)$ where $b\in [-\frac{1}{2},\frac{1}{2}]$ in a single-state SQKD protocol in Ref. \cite{21}. $U_{F}$ and $U_{R}$ denote the attack operators performed by Eve in the forward  and reverse channel respectively. $U$ is an unitary operator acting on the joint system $\mathcal{H}_{T}\otimes\mathcal{H}_{E}$. We describe Eve's restriction attack strategy as follows:
\begin{enumerate}
\item Alice prepares and sends state $|+\rangle$ to Bob through the forward channel. Eve intercepts $|+\rangle$ and resends another state $|e\rangle$ prepared by herself to Bob, where

\begin{eqnarray}
&& |e\rangle=\sqrt{\frac{1}{2}+b}|0\rangle+\sqrt{\frac{1}{2}-b}|1\rangle.
\end{eqnarray}

\item Bob has two choices when he receives the state $|e\rangle$.

Case 1: Bob chooses to CTRL the state, then he reflects it undisturbed to Alice through the reverse channel. Meanwhile, Eve will capture the transit state and probe it using unitary operator $U$ acting on the transit state and entangling it with her own ancilla state. After that Eve resends the transit state to Alice and keeps the ancilla state in her own memory.

Case 2: Bob chooses to SIFT the state, then Bob will send $|0\rangle$ and $|1\rangle$ to Alice with probabilities $\frac{1}{2}+b$ and $\frac{1}{2}-b$ respectively. Eve can also perform the attack like Case 1 during the transmission.

\end{enumerate}

$b$ is a bias parameter, it can be observed by the legitimate users.
Eve probes the state by using a unitary operator $U$ to act on $\mathcal{H}_{T}\otimes \mathcal{H}_{E}$ as follows:
\begin{eqnarray}
&& U|0,0\rangle = |0,e_{00}\rangle+|1,e_{01}\rangle,\\
&& U|1,0\rangle = |0,e_{10}\rangle+|1,e_{11}\rangle.
\end{eqnarray}

Since $U$ is a unitary operator, we have
\begin{eqnarray}
&& \langle e_{00}|e_{10}\rangle + \langle e_{01}|e_{11}\rangle = 0,\\
&& \langle e_{00}|e_{00}\rangle + \langle e_{01}|e_{01}\rangle =  \langle e_{10}|e_{10}\rangle + \langle e_{11}|e_{11}\rangle =1.
\end{eqnarray}
In order to illustrate Eve's attack under the circumstance Bob chooses to CTRL the state and Alice chooses to measure in $X$ basis, we express $|e\rangle$ in $X$ basis as
\begin{eqnarray}
&& |e\rangle = \frac{\alpha+\beta}{\sqrt{2}} |+\rangle + \frac{\alpha-\beta}{\sqrt{2}} |-\rangle, \\
&& \alpha = \sqrt{\frac{1}{2}+b} ,  \beta = \sqrt{\frac{1}{2}-b}.
\end{eqnarray}
According to Eqs. (11) and (12), we can get
\begin{eqnarray}
&& U|+,0\rangle = |+,f_{+0}\rangle+ |-,f_{+1}\rangle, \\
&& U|-,0\rangle = |+,f_{-0}\rangle+ |-,f_{-1}\rangle,\\
&& |f_{+0}\rangle = \frac{1}{2}(|e_{00}\rangle+ |e_{01}\rangle+|e_{10}\rangle+|e_{11}\rangle), \\
&& |f_{+1}\rangle = \frac{1}{2}(|e_{00}\rangle- |e_{01}\rangle+|e_{10}\rangle-|e_{11}\rangle), \\
&& |f_{-0}\rangle = \frac{1}{2}(|e_{00}\rangle+ |e_{01}\rangle-|e_{10}\rangle-|e_{11}\rangle), \\
&& |f_{-1}\rangle = \frac{1}{2}(|e_{00}\rangle- |e_{01}\rangle-|e_{10}\rangle+|e_{11}\rangle).
\end{eqnarray}
Then we can get
\begin{eqnarray}
&& U|e,0\rangle = |+,g_{+}\rangle+ |-,g_{-}\rangle, \\
&& |g_{+}\rangle = \frac{\alpha}{\sqrt{2}}|e_{00}\rangle+ \frac{\alpha}{\sqrt{2}}|e_{01}\rangle+ \frac{\beta}{\sqrt{2}}|e_{10}\rangle +\frac{\beta}{\sqrt{2}}|e_{11}\rangle,\\
&& |g_{-}\rangle = \frac{\alpha}{\sqrt{2}}|e_{00}\rangle- \frac{\alpha}{\sqrt{2}}|e_{01}\rangle+ \frac{\beta}{\sqrt{2}}|e_{10}\rangle- \frac{\beta}{\sqrt{2}}|e_{11}\rangle.
\end{eqnarray}

Next, we model one iteration of this protocol as follows:

\begin{enumerate}

\item Alice prepares and sends the state $|+\rangle$ to Bob through the forward channel:
\begin{eqnarray}
&& \rho_{1} = |+\rangle\langle+|_{T}.
\end{eqnarray}

\item Eve performs the restricted operation $U_{F}$ on the transit state. Accordingly, it is equivalent to intercept the state $|+\rangle$ and resend $|e\rangle$ to Bob:
\begin{eqnarray}
&& \rho_{2} = U_{F}|+\rangle\langle+|_{T}U^{*}_{F} = |e\rangle\langle e|_{T}.
\end{eqnarray}

\item Bob's action:

(1) SIFT:

\begin{eqnarray}
&& \rho^{S}_{3} = |0\rangle\langle0|_{B}\otimes \alpha^{2}|0\rangle\langle0|_{T}+|1\rangle\langle1|_{B}\otimes \beta^{2}|1\rangle\langle1|_{T}.
\end{eqnarray}

(2) CTRL:

\begin{eqnarray}
&& \rho^{C}_{3} = |e\rangle\langle e|_{T}.
\end{eqnarray}

\item Eve attacks in the reverse channel:

(1) SIFT:

\begin{eqnarray}
&& \rho^{S}_{4} = |0\rangle\langle0|_{B}\otimes \sigma_{1} +|1\rangle\langle1|_{B}\otimes \sigma_{2}, \\
&& \sigma_{1}=\alpha^{2}P(|0,e_{00}\rangle_{TE}+|1,e_{01}\rangle_{TE}),\\
&& \sigma_{2}=\beta^{2}P(|0,e_{10}\rangle_{TE}+|1,e_{11}\rangle_{TE}),\\
&& P(|x\rangle)=|x\rangle \langle x|.
\end{eqnarray}

(2) CTRL:

\begin{eqnarray}
&& \rho^{C}_{4} = U|e,0\rangle\langle e,0|_{TE}U^{*}=P(|+,g_{+}\rangle_{TE}+ |-,g_{-}\rangle_{TE}).\\
\end{eqnarray}

\item Alice measures in $Z$ or $X$ basis randomly:

(1) SIFT-Z :

\begin{eqnarray}
&&\rho^{S-Z}_{5} = |0\rangle\langle0|_{A}\otimes(|0\rangle\langle0|_{B}\otimes \alpha^{2}|e_{00}\rangle\langle e_{00}|_{E}+|1\rangle\langle1|_{B}\otimes \beta^{2}|e_{10}\rangle\langle e_{10}|_{E})\\\nonumber
&& \quad \quad \quad+|1\rangle\langle1|_{A}\otimes(|0\rangle\langle0|_{B}\otimes \alpha^{2}|e_{01}\rangle\langle e_{01}|_{E}+|1\rangle\langle1|_{B}\otimes \beta^{2}|e_{11}\rangle\langle e_{11}|_{E}).
\end{eqnarray}

(2) SIFT-X :

\begin{eqnarray}
&&\rho^{S-X}_{5} =|+\rangle\langle+|_{A}\otimes[|0\rangle\langle0|_{B}\otimes\frac{\alpha^{2}}{2}\sigma_{+0}+|1\rangle\langle1|_{B}\otimes\frac{\beta^{2}}{2}\sigma_{+1}]\\\nonumber
&&\quad \quad \quad +|-\rangle\langle-|_{A}\otimes[|0\rangle\langle0|_{B}\otimes\frac{\alpha^{2}}{2}\sigma_{-0}+|1\rangle\langle1|_{B}\otimes\frac{\beta^{2}}{2}\sigma_{-1}],\\
&& \sigma_{+0}= |e_{00}\rangle\langle e_{00}|_{E}+|e_{00}\rangle\langle e_{01}|_{E}+|e_{01}\rangle\langle e_{00}|_{E}+|e_{01}\rangle\langle e_{01}|_{E},\\
&& \sigma_{+1}= |e_{10}\rangle\langle e_{10}|_{E}+|e_{10}\rangle\langle e_{11}|_{E}+|e_{11}\rangle\langle e_{10}|_{E}+|e_{11}\rangle\langle e_{11}|_{E},\\
&& \sigma_{-0}= |e_{00}\rangle\langle e_{00}|_{E}- |e_{00}\rangle\langle e_{01}|_{E}-|e_{01}\rangle\langle e_{00}|_{E}+|e_{01}\rangle\langle e_{01}|_{E},\\
&& \sigma_{-1}= |e_{10}\rangle\langle e_{10}|_{E}- |e_{10}\rangle\langle e_{11}|_{E}-|e_{11}\rangle\langle e_{10}|_{E}+|e_{11}\rangle\langle e_{11}|_{E}.
\end{eqnarray}

(3) CTRL-X :

\begin{eqnarray}
\rho^{C-X}_{5} = |+\rangle\langle+|_{A}\otimes |e\rangle\langle e|_{T} \otimes |g_{+}\rangle\langle g_{+}|_{E}+|-\rangle\langle-|_{A}\otimes |e\rangle\langle e|_{T} \otimes |g_{-}\rangle\langle g_{-}|_{E}.
\end{eqnarray}

(4) CTRL-Z :

\begin{eqnarray}
&& \rho^{C-Z}_{5} = |0\rangle\langle0|_{A}\otimes |e\rangle\langle e|_{B} \otimes \sigma_{0e}+ |1\rangle\langle 1|_{A}\otimes |e\rangle\langle e|_{B} \otimes \sigma_{1e},\\
&& \sigma_{0e}=\frac{1}{2}(|g_{+}\rangle\langle g_{+}|_{E}+|g_{+}\rangle\langle g_{-}|_{E}+|g_{-}\rangle\langle g_{+}|_{E}+|g_{-}\rangle\langle g_{-}|_{E}),\\
&& \sigma_{1e}=\frac{1}{2}(|g_{+}\rangle\langle g_{+}|_{E}-|g_{+}\rangle\langle g_{-}|_{E}-|g_{-}\rangle\langle g_{+}|_{E}+|g_{-}\rangle\langle g_{-}|_{E}).
\end{eqnarray}

\end{enumerate}

From the protocol we can see only the SIFT-$Z$ bits contribute to the raw key, we have modeled the process that Bob chooses to SIFT the state and Alice chooses to measure in the $Z$ basis as Eq. (36). We define $P(i,j), i,j \in \{0,1\}$ to denote the probability of the event that Alice and Bob's raw key bit are $i$ and $j$ respectively, and then we can get
\begin{eqnarray}
&& P(0,0)=tr[(|0,0\rangle\langle 0,0|_{AB}\otimes I)\rho^{S-Z}_{5}]=\alpha^{2}\langle e_{00}|e_{00}\rangle=(\frac{1}{2}+b)\langle e_{00}|e_{00}\rangle,\\
&& P(0,1)=tr[(|0,1\rangle\langle 0,1|_{AB}\otimes I)\rho^{S-Z}_{5}]=\beta^{2}\langle e_{10}|e_{10}\rangle=(\frac{1}{2}-b)\langle e_{10}|e_{10}\rangle,\\
&& P(1,0)=tr[(|1,0\rangle\langle 1,0|_{AB}\otimes I)\rho^{S-Z}_{5}]=\alpha^{2}\langle e_{01}|e_{01}\rangle=(\frac{1}{2}+b)\langle e_{01}|e_{01}\rangle,\\
&& P(1,1)=tr[(|1,1\rangle\langle 1,1|_{AB}\otimes I)\rho^{S-Z}_{5}]=\beta^{2}\langle e_{11}|e_{11}\rangle=(\frac{1}{2}-b)\langle e_{11}|e_{11}\rangle.
\end{eqnarray}

\subsection{Bounding the final key rate}

Considering only the SIFT-Z bits can contribute to the raw key, then the state of the joint system after an iteration of the protocol is

\begin{eqnarray}
&&\rho_{ABE} = |0\rangle\langle0|_{A}\otimes(|0\rangle\langle0|_{B}\otimes \alpha^{2}|e_{00}\rangle\langle e_{00}|_{E}+|1\rangle\langle1|_{B}\otimes \beta^{2}|e_{10}\rangle\langle e_{10}|_{E})\\\nonumber
&& \quad \quad \quad +|1\rangle\langle1|_{A}\otimes(|0\rangle\langle0|_{B}\otimes \alpha^{2}|e_{01}\rangle\langle e_{01}|_{E}+|1\rangle\langle1|_{B}\otimes \beta^{2}|e_{11}\rangle\langle e_{11}|_{E}).
\end{eqnarray}
According to Eq. (9),  inf $(S(B|E)-H(B|A))$ is a lower bound of the key rate $r$. Due to the strong subadditivity of von Neumann entropy expressed as

\begin{eqnarray}
&& S(B|E)\geq S(B|ME),
\end{eqnarray}
we can get
\begin{eqnarray}
r=\lim_{N\rightarrow\infty}\frac{\ell(N)}{N}\geq\inf(S(B|E)-H(B|A)) \geq \inf(S(B|ME)-H(B|A)),
\end{eqnarray}
where $M$ is a new system introduced into to form a compound system $ABME$. Then we introduce a new system $M$ modeled by a two-dimensional Hilbert space spanned by the orthonormal basis $\{|0\rangle, |1\rangle\}$. We use the operator $|i\rangle\langle i|_{M}, i\in\{0,1\}$ to register the $xor$ operation of $A$ and $B$'s raw key bit. We express the mixed state of the system $ABME$ as

\begin{eqnarray}
&& \rho_{ABME}=  |0\rangle\langle 0|_{A}\otimes |0\rangle\langle 0|_{B} \otimes |0\rangle\langle 0|_{M} \otimes \alpha^{2}|e_{00}\rangle\langle e_{00}|_{E}\\\nonumber
&& \quad \quad \quad \quad   +|0\rangle\langle 0|_{A}\otimes |1\rangle\langle 1|_{B} \otimes |1\rangle\langle 1|_{M} \otimes \beta^{2}|e_{10}\rangle\langle e_{10}|_{E}\\\nonumber
&& \quad \quad \quad \quad   + |1\rangle\langle 1|_{A}\otimes |0\rangle\langle 0|_{B} \otimes |1\rangle\langle 1|_{M} \otimes \alpha^{2}|e_{01}\rangle\langle e_{01}|_{E}\\\nonumber
&& \quad \quad \quad \quad   + |1\rangle\langle 1|_{A}\otimes |1\rangle\langle 1|_{B} \otimes |0\rangle\langle 0|_{M} \otimes \beta^{2}|e_{11}\rangle\langle e_{11}|_{E}.
\end{eqnarray}
Tracing out the system $A$, we can get the state $\rho_{BME}$ as

\begin{eqnarray}
\rho_{BME}=   |0\rangle\langle 0|_{B} \otimes [|0\rangle\langle 0|_{M} \otimes \alpha^{2}|e_{00}\rangle\langle e_{00}|_{E}+|1\rangle\langle 1|_{M} \otimes \alpha^{2}|e_{01}\rangle\langle e_{01}|_{E}]\\\nonumber
\quad \quad \quad \quad   +|1\rangle\langle 1|_{B} \otimes [|1\rangle\langle 1|_{M} \otimes \beta^{2}|e_{10}\rangle\langle e_{10}|_{E}+|0\rangle\langle 0|_{M} \otimes \beta^{2}|e_{11}\rangle\langle e_{11}|_{E}].
\end{eqnarray}
Then we get $\rho_{ME}$ as
\begin{eqnarray}
&& \rho_{ME} =tr_{B}(\rho_{BME})= k_{1}|0\rangle\langle 0|_{M} \otimes \rho^{0}_{E}+k_{2}|1\rangle\langle 1|_{M} \otimes \rho^{1}_{E}, \\
&&  \quad\quad k_{1}=P(0,0)+P(1,1),\\
&&  \quad\quad k_{2}=P(0,1)+P(1,0),\\
&& \quad\quad\rho^{0}_{E} = \frac{\alpha^{2}|e_{00}\rangle\langle e_{00}|_{E}+\beta^{2}|e_{11}\rangle\langle e_{11}|_{E}}{P(0,0)+P(1,1)},\\
&& \quad\quad \rho^{1}_{E} = \frac{\alpha^{2}|e_{01}\rangle\langle e_{01}|_{E}+ \beta^{2}|e_{10}\rangle\langle e_{10}|_{E}}{P(0,1)+P(1,0)}.
\end{eqnarray}

Next, we need to compute the von Neumann entropies of the system $BME$ and system $ME$. In order to compute $S(\rho_{BME})$, we rewrite $\rho_{BME}$ as a classical quantum state as
\begin{eqnarray}
&& \rho_{BME}= P(0,0)|0,0\rangle\langle 0,0|_{BM} \otimes \rho^{00}_{E}+P(0,1)|1,1\rangle\langle 1,1|_{BM}  \otimes \rho^{01}_{E}\\\nonumber
&& \quad \quad\quad  +P(1,0)|0,1\rangle\langle 0,1|_{BM} \otimes \rho^{10}_{E}+P(1,1)|1,0\rangle\langle 1,0|_{BM} \otimes \rho^{11}_{E},\\
&&   \rho^{00}_{E}=\frac{|e_{00}\rangle\langle e_{00}|_{E}}{\langle  e_{00}| e_{00} \rangle},\quad \quad \rho^{01}_{E}=\frac{|e_{10}\rangle\langle e_{10}|_{E}}{\langle  e_{10}| e_{10} \rangle},\\
&& \rho^{10}_{E}=\frac{|e_{01}\rangle\langle e_{01}|_{E}}{\langle  e_{01}| e_{01} \rangle},\quad \quad \rho^{11}_{E}=\frac{|e_{11}\rangle\langle e_{11}|_{E}}{\langle  e_{11}| e_{11} \rangle}.
\end{eqnarray}
Because $\rho^{ij}_{E}, i,j \in \{0,1\}$ are all pure state,  $S(\rho^{ij}_{E})=0$. According to Eq. (7), we can figure out
\begin{eqnarray}
&&S(BME)=S(\rho_{BME})=H(P(0,0),P(0,1),P(1,0),P(1,1)).
\end{eqnarray}
Next, we compute $S(ME)$ as
\begin{eqnarray}
&&S(ME)=k_{0}+k_{1}S(\rho^{0}_{E})+k_{2}S(\rho^{1}_{E}),\\
&& k_{0}=h(P(0,0)+P(1,1)).
\end{eqnarray}
As $k_{2}=P(0,1)+P(1,0)$ is considered as the error rate in the SIFT-Z bits,  it should be very small; otherwise, the protocol will be aborted. $\rho^{1}_{E}$ is a two-dimensional density operator, then
\begin{eqnarray}
&& S(\rho^{1}_{E})\leq 1.
\end{eqnarray}
Therefore, we can bound $S(ME)$ as
\begin{eqnarray}
&& S(ME)=k_{0}+k_{1}S(\rho^{0}_{E})+k_{2}S(\rho^{1}_{E})\leq k_{0}+k_{2}+k_{1}S(\rho^{0}_{E}).
\end{eqnarray}
According to Eqs. (8)(52)(63)(67), we can get
\begin{eqnarray}
&& r\geq H(P(i,j)_{ij})-k_{0}-k_{2}-k_{1}S(\rho^{0}_{E})-H(B|A).
\end{eqnarray}
Then we can compute a lower bound of the key rate $r$ by finding an upper bound on $S(\rho^{0}_{E})$.

Let $\alpha|e_{00}\rangle= x |\xi\rangle$ and $\beta|e_{11}\rangle= y |\xi\rangle + z |\eta\rangle$, where $x,y,z\in\mathbb{C}$, $\langle\xi|\xi\rangle=\langle\eta|\eta\rangle=1$ and $\langle\xi|\eta\rangle=0$. Then we can get
\begin{eqnarray}
&&|x|^{2}=\alpha^{2}\langle e_{00}| e_{00}\rangle = P(0,0),\\
&&|y|^{2}+|z|^{2}=\beta^{2}\langle e_{11}| e_{11}\rangle = P(1,1),\\
&& x^{*}y = \alpha\beta \langle e_{00}| e_{11}\rangle,\\
&&|y|^{2}=\frac{\alpha^{2}\beta^{2}|\langle e_{00}|e_{11}\rangle|^{2}}{|x|^{2}}.
\end{eqnarray}
In the basis of $\{|\xi\rangle, |\eta\rangle\}$, we can write $\rho^{0}_{E}$ as
\begin{equation}
\rho^{0}_{E}=\frac{1}{|x|^{2}+|y|^{2}+|z|^{2}}
\left(                 
  \begin{array}{cc}   
    |x|^{2}+|y|^{2}, & \quad yz^{*} \\  
    \quad y^{*}z \quad\quad , & \quad|z|^{2} \\  
  \end{array}
\right).                 
\end{equation}
Its eigenvalues are
\begin{eqnarray}
&& \lambda_{\pm}=\frac{1}{2}\pm\frac{\sqrt{|x|^{4}+|y|^{4}+|z|^{4}+2|x|^{2}|y|^{2}+2|y|^{2}|z|^{2}-2|x|^{2}|z|^{2}}}{2(|x|^{2}+|y|^{2}+|z|^{2})}.
\end{eqnarray}
Then we can get
\begin{eqnarray}
&& \lambda_{\pm}=\frac{1}{2}\pm\frac{\sqrt{(P(0,0)-P(1,1))^{2}+4\alpha^{2}\beta^{2}|\langle e_{00}|e_{11}\rangle|^{2}}}{2(P(0,0)+P(1,1))}
\end{eqnarray}
according to Eqs. (65), (66), (67) and (68).
Thus, we have
\begin{eqnarray}
&& S(\rho^{0}_{E})=h(\lambda_{+})
\end{eqnarray}
which is a bound for $\alpha^{2}\beta^{2}|\langle e_{00}|e_{11}\rangle|^{2}$. From Eq. (75), we can see $\lambda_{+}\geq\frac{1}{2}$, and then as $\lambda_{+}$ decreases, $h(\lambda_{+})$ increases. We can use $\alpha^{2}\beta^{2}|\langle e_{00}|e_{11}\rangle|^{2}$'s lower bound to find an upper bound of $S(\rho^{0}_{E})$. Assume $\mathcal{B}\geq0$ is a lower bound of $\alpha\beta|\langle e_{00}|e_{11}\rangle|$ and define
\begin{eqnarray}
&& \lambda=\frac{1}{2}+\frac{\sqrt{(P(0,0)-P(1,1))^{2}+4\mathcal{B}^{2}}}{2(P(0,0)+P(1,1))},
\end{eqnarray}
then have
\begin{eqnarray}
&& S(ME)\leq k_{0}+k_{2}+k_{1}h(\lambda).
\end{eqnarray}

Next, we compute $H(B|A)$ by the observable statistics $P(i,j)_{ij}$. We can easily get
\begin{eqnarray}
&& H(BA)=H(\{P(i,j)\}_{ij}).
\end{eqnarray}
We can get
\begin{eqnarray}
&& P_{A}(0)=P(0,0)+P(0,1),\\
&& P_{A}(1)=P(1,0)+P(1,1),
\end{eqnarray}
and,
\begin{eqnarray}
&& H(A)=h(P_{A}(0))=h(P(0,0)+P(0,1)).
\end{eqnarray}
Thus,
\begin{eqnarray}
&& \quad H(B|A)=H(BA)-H(A)=H(\{P(i,j)\}_{ij})-h(P(0,0)+P(0,1)).
\end{eqnarray}
Therefore, we can bound the final key rate as
\begin{eqnarray}
r\geq h(P(0,0)+P(0,1))-k_{0}-k_{2}-k_{1}h(\lambda).
\end{eqnarray}
From the above inequality, we can see all the parameters can be estimated by $A$ and $B$ except $\alpha\beta|\langle e_{00}|e_{11}\rangle|$'s lower bound $\mathcal{B}$. Next, we also consider to use some other observable statistics to determine a value of $\mathcal{B}$.

\subsection{Bounding $\alpha\beta|\langle e_{00}|e_{11}\rangle|$ using observable statistics}
In the previous subsection, we mainly concern with the SIFT-Z process of the protocol. Now we begin to talk on some other ones by using some mismatched measurements \cite{32,33,34,35,36}.

Firstly, we consider the process of CTRL-X. According to the protocol, we will abort it when there is too much noise in the quantum communication channel. Then the legitimate users can estimate noise of the X-basis type in this step. Specifically, if Bob chooses to CTRL the state and Alice chooses to measure in X basis and gets the outcome $|-\rangle$, it is considered as a mistake because of the channel noise. We use $P(e,-)$ to denote the probability of the event Alice gets $-$ in the CTRL-X process, we can compute it as
\begin{eqnarray}
&& P(e,-)=tr[(|-\rangle\langle-|_{A}\otimes|e\rangle\langle e|_{T}\otimes I)\rho^{C-X}_{5}]=\langle g_{-}|g_{-}\rangle=\frac{1}{2}- k_{3}.\\
&& k_{3}=\alpha^{2}Re\langle e_{00}|e_{01}\rangle+\alpha\beta Re\langle e_{00}|e_{11}\rangle+\alpha\beta Re\langle e_{01}|e_{10}\rangle+\beta^{2} Re\langle e_{10}|e_{11}\rangle.
\end{eqnarray}
Thus we can specify $\alpha\beta Re\langle e_{00}|e_{11}\rangle$  as
\begin{eqnarray}
&& \alpha\beta Re\langle e_{00}|e_{11}\rangle=\frac{1}{2}-P(e,-)-k_{4}.\\
&& k_{4}=\alpha^{2}Re\langle e_{00}|e_{01}\rangle+\alpha\beta Re\langle e_{01}|e_{10}\rangle+\beta^{2} Re\langle e_{10}|e_{11}\rangle.
\end{eqnarray}
From Eqs. (87) and (88), we can see the right side of the Eq. (87) contains the parameters $\alpha^{2}Re\langle e_{00}|e_{01}\rangle,\alpha\beta Re\langle e_{01}|e_{10}\rangle$ and $\beta^{2} Re\langle e_{10}|e_{11}\rangle$.
Next, we use the observable statistics to specify them one by one.
\begin{enumerate}
\item  $\alpha^{2}Re\langle e_{00}|e_{01}\rangle$ :

Let $P(0,+)$ denote the probability of the event Alice gets the outcome $+$ when Bob resends $|0\rangle$ in the SIFT-X process. Then we can get
\begin{eqnarray}
&& P(0,+)=tr[(|+\rangle\langle+|_{A}\otimes|0\rangle\langle 0|_{T}\otimes I)\rho^{S-X}_{5}]= \frac{\alpha^{2}}{2}+\alpha^{2}Re\langle e_{00}|e_{01}\rangle.
\end{eqnarray}
Then $\alpha^{2}Re\langle e_{00}|e_{01}\rangle$ can be specified by
\begin{eqnarray}
&& \alpha^{2} Re\langle e_{00}|e_{01}\rangle = P(0,+)-\frac{\alpha^{2}}{2}.
\end{eqnarray}

\item $ \beta^{2} Re\langle e_{10}|e_{11}\rangle$ :

Let $P(1,+)$ denote the probability of the event Alice get the outcome $+$ when Bob resends $|1\rangle$ in the SIFT-X process. Then we can get
\begin{eqnarray}
&& P(1,+)=tr[(|+\rangle\langle+|_{A}\otimes|1\rangle\langle 1|_{T}\otimes I)\rho^{S-X}_{5}]= \frac{\beta^{2}}{2}+\beta^{2}Re\langle e_{10}|e_{11}\rangle.
\end{eqnarray}
Thus,
\begin{eqnarray}
&& \beta^{2} Re\langle e_{10}|e_{11}\rangle =P(1,+)-\frac{ \beta^{2} }{2}.
\end{eqnarray}

\item  $\alpha\beta Re\langle e_{01}|e_{10}\rangle$ :

Here we cannot specify $\alpha\beta Re\langle e_{01}|e_{10}\rangle$ exactly, but we can bound it using the Cauchy-Schwarz inequality as
\begin{eqnarray}
&& \alpha\beta Re\langle e_{01}|e_{10}\rangle \leq  \sqrt{\alpha^{2}\beta^{2}\langle e_{01}|e_{01}\rangle\langle e_{10}|e_{10}\rangle}=\sqrt{P(0,1)P(1,0)}.
\end{eqnarray}
\end{enumerate}

Then we can bound $\alpha\beta\langle e_{00}|e_{11}\rangle$ as
\begin{eqnarray}
&& \alpha\beta Re\langle e_{00}|e_{11}\rangle \geq  1-P(e,-)-P(0,+)-P(1,+)-\sqrt{P(0,1)P(1,0)}.
\end{eqnarray}
Here we define $\mathcal{B}$ as
\begin{eqnarray}
&& \mathcal{B}= 1-P(e,-)-P(0,+)-P(1,+)-\sqrt{P(0,1)P(1,0)}.
\end{eqnarray}
Then we can ensure $\mathcal{B}> 0$ by controlling the quantum channel noise because all the observable statistics are determined by the quantum channel from Bob to Alice. If the channel is too noisy, it will be  aborted definitely. Thus, we can get the lower bound on $\alpha\beta|\langle e_{00}|e_{11}\rangle|$ because of
\begin{eqnarray}
&& \alpha\beta|\langle e_{00}|e_{11}\rangle|=\alpha\beta\sqrt{Re^{2}(\langle e_{00}|e_{11}\rangle)+Im^{2}(\langle e_{00}|e_{11}\rangle)}\\\nonumber
&&\quad \quad \quad \quad \quad \quad \geq \alpha\beta |Re\langle e_{00}|e_{11}\rangle| \geq \alpha\beta Re\langle e_{00}|e_{11}\rangle \geq \mathcal{B}.
\end{eqnarray}

From above, we have got a lower bound of the key rate $r$, which is expressed as a function of parameters determined by the quantum channel. According to this bound, we can compute a threshold value such that the secret keys can be established successfully as long as all the error rates are below this value. Then the security proof restricted on Eve's collective attack is finished.

\subsection{Security on the circumstance of general attack}
In order to get the full unconditional security proof, we need to spread it to the circumstance that Eve can perform the most powerful attack strategy - general attack. $Renner$ et al \cite{26} have pointed out that it suffices to consider the collective attacks when analyzing the full security of the QKD protocols which are with the character of permutation invariant. $Renner$ et al  \cite{26} have also analyzed the unconditional security of $BB84$, $B92$ and six-state QKD protocols with one-way error correction and privacy amplification by using this method. Here we declare that all above hold still in the case of general attack since the protocol is also permutation invariant. Though this SQKD protocol relies on the two-way quantum channel, $Krawec$ \cite{21} has proved that all attacks in the forward channel are equivalent to a restricted attack. That is to say, Alice sending $|+\rangle$ to Bob through the forward channel is equal to Bob preparing a state $|e\rangle$. Then the SQKD protocol can be reduced to a fully QKD protocol that Bob prepares and sends a state of a set $\{|0\rangle, |1\rangle,|e\rangle\}$ to Alice with a certain probability, Alice chooses to perform a $Z$-basis or $X$-basis measurement randomly when she receives a state from Bob. Additionally, only the SIFT-Z bits contribute to the raw key in the SQKD protocol. It is indicated that Bob's sending $|e\rangle$ to Alice is only to check the channel noise in the reduced QKD protocol. Thus, the reduced QKD protocol can be considered as part of $BB84$ protocol with an additional channel noise checking process. Then we can infer that the SQKD protocol is permutation invariant as well. The full security analysis is completed.

\section{An Example}
Here we demonstrate how to compute a lower bound of the final key rate $r$ in asymptotic scenario under the circumstance that the reverse channel is a depolarizing one with parameter $q$. The depolarization channel is a typical scenario considered in the unconditional security proofs of some other protocols \cite{19,37,38}. It acts on two-dimensional density operators $\rho$ as follows:

\begin{eqnarray}
&& \xi_{q}(\rho)=(1-q)\rho+\frac{q}{2}I,
\end{eqnarray}
where $I$ is the identity operator.

We model Eve's attack  in the reverse channel of the protocol briefly under this circumstance as follows:
\begin{enumerate}
\item SIFT :
\begin{eqnarray}
&& \rho_{S}=\alpha^{2}|0\rangle\langle 0|_{B}\otimes \xi_{q}(|0\rangle\langle 0|_{T})+\beta^{2}|1\rangle\langle 1|_{B}\otimes \xi_{q}(|1\rangle\langle 1|_{T})\\\nonumber
&& \quad =(1-\frac{q}{2})\alpha^{2}|0,0\rangle\langle 0,0|_{BT}+\frac{q}{2}\alpha^{2}|0,1\rangle\langle 0,1|_{BT}\\\nonumber
&& \quad +\frac{q}{2}\beta^{2}|1,0\rangle\langle 1,0|_{BT}+(1-\frac{q}{2})\beta^{2}|1,1\rangle\langle 1,1|_{BT}.
\end{eqnarray}

\item CTRL :
\begin{eqnarray}
&& \rho_{C}=\xi_{q}(|e\rangle\langle e|_{T})=(1-q)|e\rangle\langle e|_{T}+\frac{q}{2}(|e\rangle\langle e|_{T}+|e^{\bot}\rangle\langle e^{\bot}|_{T}),
\end{eqnarray}
where $|e^{\bot}\rangle$ is a state orthogonal to $|e\rangle$, that is to say,
\begin{eqnarray}
|e^{\bot}\rangle=\sqrt{\frac{1}{2}-b}|0\rangle-\sqrt{\frac{1}{2}+b}|1\rangle.
\end{eqnarray}

\end{enumerate}
Then we can get
\begin{eqnarray}
 && P(0,+)=tr[(|0,+\rangle\langle0,+|_{BT})\rho_{S}]=\frac{\alpha^{2}}{2},\\
 && P(1,+)=tr[(|1,+\rangle\langle1,+|_{BT})\rho_{S}]=\frac{\beta^{2}}{2},\\
 && P(e,-)=tr[(|-\rangle\langle-|_{T})\rho_{C}]=\frac{1}{2}+\frac{(q-1)}{2}\sqrt{1-4b^{2}},\\
 && P(0,0)=tr[(|0,0\rangle\langle0,0|_{BT})\rho_{S}]= (\frac{1}{2}+b)(1-\frac{q}{2}),\\
 && P(0,1)=tr[(|1,0\rangle\langle1,0|_{BT})\rho_{S}]= (\frac{1}{2}-b)\frac{q}{2},\\
 && P(1,0)=tr[(|0,1\rangle\langle0,1|_{BT})\rho_{S}]= (\frac{1}{2}+b)\frac{q}{2},\\
 && P(1,1)=tr[(|1,1\rangle\langle1,1|_{BT})\rho_{S}]= (\frac{1}{2}-b)(1-\frac{q}{2}).
\end{eqnarray}

Thus, we can get $\mathcal{B}$ as
\begin{eqnarray}
&& \mathcal{B}=(\frac{1}{2}-\frac{3q}{4})\sqrt{1-4b^{2}}.
\end{eqnarray}
Note that we can do some efforts to restrict the channel parameters $b$ and $q$ to ensure $\mathcal{B}$ is positive. Here, we only need to assume $q<\frac{2}{3}$. Then $\mathcal{B}$ will always be positive. This assumption is reasonable. If $q>\frac{2}{3}$, the reverse channel is too noisy to share secret keys, and then we have to abort it.
After that we can further get $\lambda$ according to Eq. (77) as
\begin{eqnarray}
&& \lambda=\frac{1}{2}+\frac{\sqrt{b^{2}(2-q)^{2}+4\mathcal{B}^{2}}}{2-q}.
\end{eqnarray}
Finally, we can get a lower bound of the key rate $r$  expressed as a function of parameters $b$ and $q$:
\begin{eqnarray}
&& r\geq f(b,q)= h(\frac{1}{2}+b-bq)-h(1-\frac{q}{2})-\frac{q}{2}-(1-\frac{q}{2})h(\lambda).
\end{eqnarray}

A graph of the  lower bound $f(b,q)$ as a function of $q$ for different values of $b$ is shown in figure 1. In the graph, we can see when $b=0$, $f(b,q)$ is positive for all $q\leq 0.193$, which means that when the error rate $Q_{Z}=\frac{q}{2}\leq 9.65\%$, the key rate will always be positive. Different values of $b$ corresponds to different threshold values. We can see when the absolute value of $b$ is far from $0$, the threshold value becomes smaller. When $|b|>0.36$, $f(b,q)$ will always be negative for arbitrary $q$, it means that the key rate $r$ cannot be guaranteed to be positive. In fact, the bias parameter $b$ can specify the amount of noise in the forward channel in some extent as it can introduce into the X-type error rate denoted by $Q_{X}$, which means Bob may observe $|-\rangle$ when Alice sends the state $|+\rangle$ through the forward channel. We can compute it as
\begin{eqnarray}
Q_{X}=tr(|-\rangle\langle -||e\rangle\langle e|)=\frac{1}{2}-\sqrt{\frac{1}{4}-b^{2}}.
\end{eqnarray}
From above, we can see the maximum error rate $Q_{X}$ in the forward channel is no more than 15.3\% to make sure the key rate $r$ is always positive.

We use Figure 2 to illustrate $f(b,q)$ as a function of the bias parameter $b$ for different values of $q$. Indeed, we can see when $q=0$, the key rate is positive for all $|b|<0.36$. But when $q>0.2$, $f(b,q)$ will always be negative. It means that the maximum error rate $Q_{Z}$ of the reverse channel is no more than 10\%. These may satisfy the fact that the SQKD protocol can tolerate more noise in the forward channel than the reverse one. However, the error rate $Q_{X}$ in the forward channel is not bound to cause an error of the raw key bit directly, maybe it is another reason that the forward channel can endure more error rate than the reverse one of this protocol. In any case, it tell us that  we need to make more effort to control the channel noise in the reverse direction than the forward when we implement this protocol in practice.

Compared with some other protocols, this protocol can tolerate more noise. As we know, $B92$ can tolerate up to $6.50\%$ , giving an optimal choice of states \cite{38}. With respect to $Krawec$'s SQKD protocol in Ref. \cite{19}, it can tolerate up to $5.36\%$ error assuming there is no noise in the forward channel. In our proof, we can see this single-state protocol can tolerate up to $9.65\%$ error in case of the forward channel existing no noise. Recently, the original SQKD protocol of Boyer et al. has been shown to support up to 11\% error rate \cite{36}, in contrast, Boyer's protocol requires Alice to prepare and send multiple states.
\begin{figure}
\center
\includegraphics[height=6cm, width=10cm]{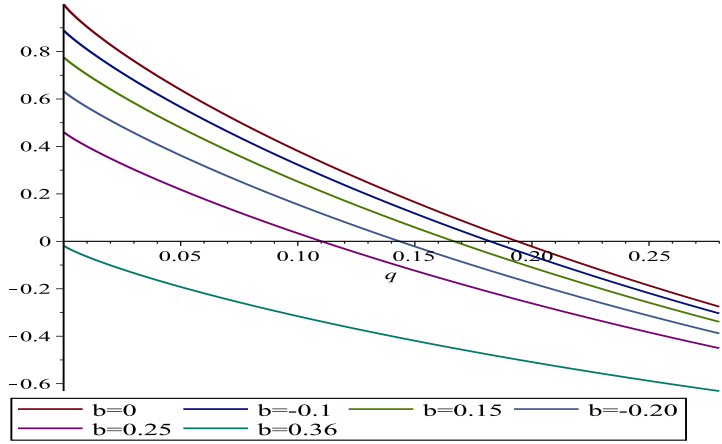}
\caption{\label{figure1}  A graph of $f(b,q)$ as a function of the depolarization channel parameter $q$ for different values of $b$. Note that, the error rate $Q_{Z}=\frac{q}{2}$.}
\end{figure}

\begin{figure}
\center
\includegraphics[height=6cm, width=10cm]{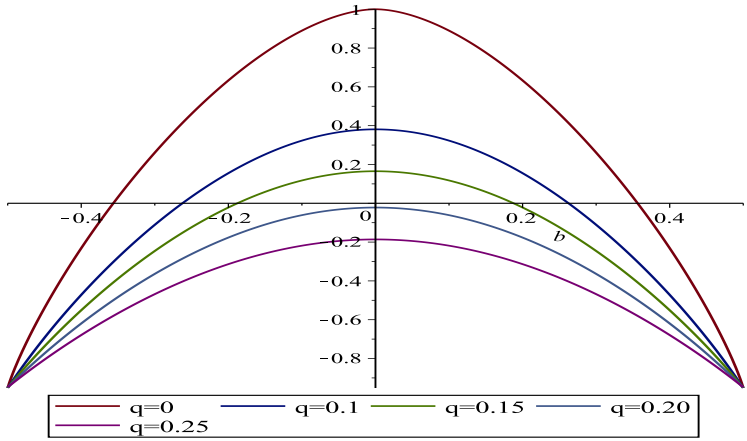}
\caption{\label{figure1}  A graph of $f(b,q)$ as a function of $b$ under the circumstance that the reverse channel is a depolarization channel of parameter $q$.}
\end{figure}

\section{Conclusion}

In this paper, we have provided an unconditional security proof of a single state SQKD protocol proposed in Ref. \cite{11} completely from information theory aspect. We have computed a lower bound of the key rate in the asymptotic scenario and derived a threshold value of errors that the protocol can tolerate. It has been showed that the secret keys can be shared successfully as long as all the channel noise in the quantum communication stage is smaller than this threshold value. Then we have made an illustration under the circumstance that the reverse channel is a depolarizing channel with parameter $q$ and showed that the SQKD protocol can sustain no more than $9.65\%$ error rate assuming that there is no noise in the forward channel. However, we have discussed the unconditional security of the protocol just under the  circumstance of perfect quantum state scenario. Therefore it is a challenge work for us to consider the unperfect scenario in the future. Meanwhile, we have only computed the key rate in the asymptotic scenario, as a consequence, the security of the protocol with finite quantum resource is another interesting problem worth considering in the future.

\begin{acknowledgements}
This work is supported in part by the National
Natural Science Foundation of China (Nos. 61272058, 61572532), the Natural Science Foundation of Qiannan Normal
College for Nationalities joint Guizhou Province of China (No. Qian-Ke-He LH Zi[2015]7719), and the Natural Science Foundation of Central Government Special Fund for Universities of West China (No. 2014ZCSX17).

\end{acknowledgements}

\end{document}